\let\orilabel\label
\documentclass[aps,pra,preprint,a4paper,showpacs,showkeys,superscriptaddress,nofootinbib]{revtex4-1}
\let\label\orilabel

\usepackage{latexsym}
\usepackage{amsmath,amssymb,mathrsfs}
\usepackage{mathtools}
\usepackage{graphicx}
\usepackage{subfigure}
\usepackage{xcolor}
\usepackage{physics}
\usepackage{cancel}
\usepackage{tensor}
\usepackage{tikz}
\usepackage{multirow,tabularx}

\newcolumntype{C}[1]{>{\centering\arraybackslash}p{#1}}
\usepackage{hyperref}     
\hypersetup{colorlinks,%
citecolor=blue,%
linkcolor=cyan,%
}
\usepackage[titletoc]{appendix}
\usepackage{enumerate}
\graphicspath{{./Figures/}}
\usetikzlibrary{decorations.pathmorphing,decorations.pathreplacing,decorations.shapes,calc}
\begin{document}


\title{Strong gravitational lensing effects of black holes with quantum hair}

\author{Sojeong Cheong}%
\email[]{jsquare@sogang.ac.kr}%
\affiliation{Department of Physics, Sogang University, Seoul, 04107, Republic of Korea}%
\affiliation{Center for Quantum Spacetime, Sogang University, Seoul, 04107, Republic of Korea}%

\author{Wontae Kim}%
\email[]{wtkim@sogang.ac.kr}%
\affiliation{Department of Physics, Sogang University, Seoul, 04107, Republic of Korea}%
\affiliation{Center for Quantum Spacetime, Sogang University, Seoul, 04107, Republic of Korea}%

\date{\today}

\begin{abstract}
	According to the no-hair theorem, stationary black holes are uniquely characterized by their mass, charge, and angular momentum.
    In this paper, we explore quantum hair by deriving the quantum-corrected black hole metric within the Barvinsky-Vilkovisky formalism.
    The quantum-corrected metric is obtained perturbatively around flat spacetime without assuming either the commutativity between the nonlocal operator and covariant derivatives or the nonlocal Gauss-Bonnet theorem, both of which are adopted in previous studies.
    Using this metric, we evaluate the deflection angle in the strong-field limit and compute the associated strong gravitational lensing observables, such as the angular separation and the relative magnification.
    Our results show that as the quantum hair, determined by the number of virtual massless quantum fields in the nonlocal effective action, increases, the photon sphere radius, the strong deflection angle, and the relative magnification all increase, whereas the angular separation decreases.
    As a result, we demonstrate that the quantum hair affects not only the black hole geometry but also its strong gravitational lensing effects.
\end{abstract}


\maketitle
\raggedbottom

\section{introduction}
\label{sec:introduction}
Classically, a stationary black hole is characterized solely by its mass, charge, and angular momentum, regardless of the nature of the collapsing matter that formed it.
This underlies the so-called no-hair theorem~\cite{Israel:1967wq,Israel:1967za,Carter:1971zc,Robinson:1975bv}, which asserts that these three quantities are all the possible distinguishing features of a black hole.
The no-hair theorem is normally well-defined in the framework of classical general relativity.
In theories that go beyond general relativity, such as those incorporating quantum gravitational effects, black holes may acquire additional degrees of freedom as \textit{quantum hair}.
It is one of considerable interests to explore the emergence of quantum hair not only in black holes but also in astrophysical objects, where similar quantum aspects may arise.

Based on the Barvinsky-Vilkovisky formalism~\cite{Barvinsky:1983vpp,Barvinsky:1985an,Barvinsky:1987uw,Barvinsky:1990up,Barvinsky:1990uq}, several studies on quantum hair have been conducted, focusing on stellar configurations~\cite{Calmet:2019eof,Calmet:2020tlj,Calmet:2021stu,Calmet:2023met,Cheong:2023oik,Perrucci:2024qrr,Cheong:2025scf}.
These studies show that the internal structure of a star, such as its mass density or internal shell distribution, plays the role of quantum hair in the quantum-corrected stellar geometry~\cite{Calmet:2021stu}; however, it is absent in the Schwarzschild black hole geometry~\cite{Calmet:2017qqa,Calmet:2018elv}.
The reason why the Schwarzschild black hole metric does not receive quantum corrections in Ref.~\cite{Calmet:2017qqa} is due to the assumption that the nonlocal operator commutes with covariant derivatives.
This assumption allows the cancellation of the nontrivial contribution from the Riemann tensor term in the equations of motion, leading to the resulting metric that remains identical to the classical one.
On the other hand, the nonlocal Gauss-Bonnet theorem can also removes the Riemann tensor term in the nonlocal effective action~\cite{Calmet:2018elv}.
Specifically, up to quadratic order in curvature, the nonlocal effective action contains the Ricci scalar squared, Ricci tensor squared, and Riemann tensor squared terms, each multiplied by a nonlocal operator,
but the last term can be rewritten in terms of the former two by the use of the nonlocal Gauss-Bonnet theorem.
Thus, there is no dependence on quantum hair in the Schwarzschild black hole metric in both Refs.~\cite{Calmet:2017qqa,Calmet:2018elv}.

In contrast to Refs.~\cite{Calmet:2017qqa,Calmet:2018elv}, assuming neither the vanishing commutation relation between the nonlocal operator and covariant derivatives nor the nonlocal Gauss-Bonnet theorem, a quantum-corrected black hole metric is derived by taking into account the nontrivial contribution of the Riemann-squared term~\cite{Xiao:2021zly}.
It is important to note, however, that the quantum-corrected black hole metric in Ref.~\cite{Xiao:2021zly} is obtained up to quadratic order in curvature from the same curvature order of the effective action.
To obtain quantum corrections up to quadratic order in curvature in the equations of motion, one should consider the effective action up to cubic order because the metric variation of an action generates terms not only at the same curvature order as the action, but also at one order lower in curvature~\cite{Cheong:2023oik}.

In this paper, we derive the quantum-corrected metric for black holes to linear order in curvature without assuming the commutativity of the nonlocal operator with covariant derivatives and the nonlocal Gauss-Bonnet theorem.
To obtain the quantum-corrected metric, we consider the metric perturbation around the flat background based on the spirit of the original Barvinsky-Vilkovisky formalism, whereas previous works carry out the metric perturbation around the Schwarzschild background~\cite{Calmet:2017qqa,Calmet:2018elv,Calmet:2019eof,Calmet:2020tlj,Calmet:2021stu,Xiao:2021zly,Cheong:2023oik,Perrucci:2024qrr,Calmet:2023met}.
In the quantum-corrected black hole metric, we explore the effect of quantum hair through strong gravitational lensing effects~\cite{Virbhadra:1999nm,Bozza:2002zj,Iyer:2006cn,Tsukamoto:2016qro,Tsukamoto:2016jzh,Soares:2025hpy}.
Considering light rays passing near the photon sphere of the quantum-corrected black hole, we investigate not only the deflection angle in the strong-field limit but also the strong gravitational lensing observables.
We numerically compute the radius of photon sphere as well as the strong-field coefficients of deflection angle, which govern the gravitational lensing phenomena in the strong-field regime.

The organization of this paper is as follows.
In Sec.~\ref{sec:metric}, we consistently derive the quantum-corrected black hole metric at linear order in curvature from the effective action up to quadratic order in curvature.
In Sec.~\ref{sec:deflections}, we introduce Bozza's method~\cite{Bozza:2002zj} to compute the deflection angle in the strong-field limit and study its dependence on quantum hair.
Quantum corrections to the strong gravitational lensing observables are then examined numerically in Sec.~\ref{sec:observables}.
In Sec.~\ref{sec:conclusion}, we present the conclusion and discussion.

\section{Quantum-corrected black hole metric}
\label{sec:metric}
We start with the bare action up to quadratic order in curvature as
\begin{equation}
	S_{\rm bare} = \int\dd[4]x \sqrt{-g} \left[ \frac{1}{16\pi G}\mathcal{R} + c_{10} \mathcal{R}^2 + c_{20}\mathcal{R}_{\mu\nu}\mathcal{R}^{\mu\nu} + c_{30}\mathcal{R}_{\mu\nu\kappa\lambda}\mathcal{R}^{\mu\nu\kappa\lambda} \right] + S_{\rm field},
\end{equation}
where $c_{10}$, $c_{20}$, and $c_{30}$ are the dimensionless bare coefficients, and $S_{\rm field}$ denotes the action for massless quantum fields including scalars, fermions, and vectors.
By integrating out the massless quantum fields in $S_{\rm field}$, the effective action up to quadratic order in curvature takes the form as~\cite{Barvinsky:1983vpp,Barvinsky:1985an,Barvinsky:1987uw,Barvinsky:1990up,Barvinsky:1990uq}
\begin{equation}
	\label{eq:eff_action}
	S_{\rm eff} = \frac{1}{16\pi G}\int\dd[4]x \sqrt{-g} \mathcal{R} + \Gamma_{\rm L} + \Gamma_{\rm NL},
\end{equation}
where $\Gamma_{\rm L}$ and $\Gamma_{\rm NL}$ are the local and nonlocal parts of the effective action, respectively.
The explicit expressions of $\Gamma_{\rm L}$ and $\Gamma_{\rm NL}$ are respectively given by
\begin{equation}
    \label{eq:local_action}
    \Gamma_{\rm L} = \int \dd[4]x \sqrt{-g} \left[ c_1(\mu) \mathcal{R}^2 + c_2(\mu) \mathcal{R}_{\mu\nu}\mathcal{R}^{\mu\nu} + c_3(\mu) \mathcal{R}_{\mu\nu\kappa\lambda}\mathcal{R}^{\mu\nu\kappa\lambda} \right]
\end{equation}
and
\begin{equation}
    \label{eq:nonlocal_action}
    \Gamma_{\rm NL}\! = \!-\! \!\int\! \dd[4]x \sqrt{-g} \!\left[ \alpha \mathcal{R}\ln\left(\! -\frac{\square}{\mu^2}\right)\mathcal{R} \!+\! \beta \mathcal{R}_{\mu\nu}\ln\left(\! -\frac{\square}{\mu^2} \right)\mathcal{R}^{\mu\nu} \!+\! \gamma \mathcal{R}_{\mu\nu\kappa\lambda}\ln\left(\! -\frac{\square}{\mu^2} \right)\mathcal{R}^{\mu\nu\kappa\lambda} \right]\! ,
\end{equation}
where $\mu$ is the renormalization scale and $\square = g^{\mu\nu}\nabla_\mu \nabla_\nu$.
In Eq.~\eqref{eq:local_action}, the local Wilson coefficients $c_1(\mu)$, $c_2(\mu)$, and $c_3(\mu)$ obey the renormalization group equations~\cite{Donoghue:2014yha,El-Menoufi:2015cqw}
\begin{align}
    c_1(\mu) &= c_1(\mu_*) - \alpha \ln \left( \frac{\mu^2}{\mu_*^2} \right),\label{eq:renormal_eq1}\\
    c_2(\mu) &= c_2(\mu_*) - \beta \ln \left( \frac{\mu^2}{\mu_*^2} \right),\label{eq:renormal_eq2}\\
    c_3(\mu) &= c_3(\mu_*) - \gamma \ln \left( \frac{\mu^2}{\mu_*^2} \right),\label{eq:renormal_eq3}
\end{align}
ensuring the scale independence of the effective action~\eqref{eq:eff_action}.
Here, $\mu_*$ is a fixed value of $\mu$ that can be determined experimentally.
In Eqs.~\eqref{eq:nonlocal_action}-\eqref{eq:renormal_eq3}, the coefficients $\alpha$, $\beta$, and $\gamma$ are the nonlocal Wilson coefficients, whose values depend on the field content of the theory.
More precisely, they are given by~\cite{Birrell:1982ix,Buchbinder:1992rb}
\begin{align}
	\alpha &= \frac{1}{11520\pi^2}\left( 5(6\xi-1)^2\mathcal{N}_{s} - 5\mathcal{N}_{f} - 50 \mathcal{N}_{v} \right),\label{eq:alpha}\\
	\beta &= \frac{1}{11520\pi^2}\left( -2\mathcal{N}_{s} +8\mathcal{N}_{f} + 176 \mathcal{N}_{v} \right),\label{eq:beta}\\
	\gamma &= \frac{1}{11520\pi^2}\left( 2\mathcal{N}_{s} + 7\mathcal{N}_{f} - 26 \mathcal{N}_{v} \right),\label{eq:gamma}
\end{align}
where $\xi$ represents the nonminimal coupling of the scalar field with the curvature, and $\mathcal{N}_{s}$, $\mathcal{N}_{f}$, and $\mathcal{N}_{v}$ are the number of scalar, fermion, and vector fields, respectively.

The metric variation of Eq.~\eqref{eq:eff_action} yields the field equations with quantum-corrected terms, written as
\begin{equation}
    \label{eq:eom}
    G_{\mu\nu} + 16\pi G (H^{\rm L}_{\mu\nu}+H^{\rm NL}_{\mu\nu}) = 0,
\end{equation}
where the Einstein tensor is $G_{\mu\nu} = \mathcal{R}_{\mu\nu} - \frac{1}{2}g_{\mu\nu}\mathcal{R}$, $H_{\mu\nu}^{\rm L} = \frac{1}{\sqrt{-g}}\fdv{\Gamma_{\rm L}}{g^{\mu\nu}}$, and $H_{\mu\nu}^{\rm NL} = \frac{1}{\sqrt{-g}}\fdv{\Gamma_{\rm NL}}{g^{\mu\nu}}$.
At linear order in curvature, the explicit forms of $H_{\mu\nu}^{\rm L}$ and $H_{\mu\nu}^{\rm NL}$ are derived as
\begin{align}
    \label{eq:local_eom}
    H^{\rm L}_{\mu\nu} &= 2\bar{c}_1 \left( g_{\mu\nu}\square \mathcal{R} - \nabla_\mu\nabla_\nu \mathcal{R} \right) + \bar{c}_2 \left( \square \mathcal{R}_{\mu\nu}  + \frac{1}{2}g_{\mu\nu}\square \mathcal{R} - \nabla_\alpha\nabla_\mu \mathcal{R}^\alpha_\nu - \nabla_\alpha\nabla_\nu \mathcal{R}^\alpha_\mu \right)
\end{align}
and
\begin{align}
    \label{eq:nonlocal_eom}
    H^{\rm NL}_{\mu\nu} =& -2\alpha \left[ g_{\mu\nu}\square - \nabla_\mu\nabla_\nu \right]\ln\left(-\frac{\square}{\mu^2}\right)\mathcal{R}\nonumber\\
    &-\beta \left[ \delta^\alpha_\mu g_{\nu\beta}\square + g_{\mu\nu}\nabla^\alpha\nabla_\beta - \delta^\alpha_\mu\nabla_\beta\nabla_\nu - \delta^\alpha_\nu\nabla_\beta\nabla_\mu \right]\ln\left(-\frac{\square}{\mu^2}\right)\mathcal{R}^\beta_\alpha\nonumber\\
    &-2\gamma \left[ \left( \delta^\alpha_\mu g_{\nu\sigma} + \delta^\alpha_\nu g_{\mu\sigma} \right) \nabla^\beta \nabla_\tau \right] \ln\left(-\frac{\square}{\mu^2}\right)\mathcal{R}^{\sigma\tau}_{\alpha\beta}.
\end{align}
In Eq.~\eqref{eq:local_eom}, we redefine the local coefficients as $\bar{c}_1=c_1-c_3$ and $\bar{c}_2=c_2+4c_3$ since the Riemann-squared term in Eq.~\eqref{eq:local_action} can be eliminated by the usual Gauss-Bonnet theorem.
Note that we do not employ the nonlocal Gauss-Bonnet theorem in Eq.~\eqref{eq:nonlocal_action}, which is different from Refs.~\cite{Calmet:2019eof,Calmet:2020tlj,Calmet:2021stu,Calmet:2023met,Cheong:2023oik,Cheong:2025scf,Calmet:2018elv}.
On the other hand, since the metric variation of the nonlocal operator in Eq.~\eqref{eq:nonlocal_action} generates terms beyond linear order in curvature~\cite{Donoghue:2014yha,Donoghue:2015nba}, we neglect those higher-order terms.

To solve the quantum-corrected field equations~\eqref{eq:eom}, we now consider the metric perturbation around flat spacetime~\cite{Barvinsky:1983vpp,Barvinsky:1985an,Barvinsky:1987uw,Barvinsky:1990up,Barvinsky:1990uq}.
Assuming the static and spherically symmetric geometry, we perturb the metric as
\begin{equation}
	\label{eq:metric perturb}
	g_{\mu\nu} = \eta_{\mu\nu} + G h_{\mu\nu}^{(1)} + G^2 h_{\mu\nu}^{(2)} + \mathcal{O}(G^3),
\end{equation}
where $\eta_{\mu\nu}\dd{x}^\mu\dd{x}^\nu = -\dd{t}^2+\dd{r}^2+r^2\left(\dd{\theta}^2+\sin^2\dd{\phi}^2\right)$ and the gravitational constant $G$ is used as the expansion parameter.
In Eq.~\eqref{eq:eom}, the $\theta\theta$ and $\phi\phi$ components in $h^{(1)}_{\mu\nu}$ and $h^{(2)}_{\mu\nu}$ can be set to zero by choosing a gauge fixing condition.
Expanding Eq.~\eqref{eq:eom} order by order in $G$, we obtain at $\mathcal{O}(G)$,
\begin{equation}
    \label{eq:eom1}
    G^{(1)}_{\mu\nu} \left[ G h^{(1)} \right] + 16\pi G \left( H^{\rm L(0)}_{\mu\nu}[\eta] + H^{\rm NL(0)}_{\mu\nu}[\eta] \right) = 0,
\end{equation}
where $G^{(1)}_{\mu\nu}$ is the linearized Einstein tensor which is written as
\begin{equation}
    G^{(1)}_{\mu\nu} [h] = \bar{\nabla}^{\alpha}\bar{\nabla}_{(\mu}h_{\nu)\alpha}- \frac{1}{2}\bar{\nabla}_{\mu}\bar{\nabla}_{\nu}h_{\alpha}^{\alpha} -\frac{1}{2}\bar{\square}h_{\mu\nu} -\frac{1}{2}\eta_{\mu\nu}\left(\bar{\nabla}^{\alpha}\bar{\nabla}^{\beta} h_{\alpha\beta} - \bar{\square}h_{\alpha}^{\alpha}\right)
\end{equation}
with the operators $\bar{\nabla}$ and $\bar{\square}$ defined with respect to the flat background.
At $\mathcal{O}(G^2)$, Eq.~\eqref{eq:eom} becomes
\begin{equation}
    \label{eq:eom2}
    G^{(1)}_{\mu\nu} \left[ G^2 h^{(2)} \right] + G^{(2)}_{\mu\nu} \left[ G h^{(1)} \right] + 16\pi G \left( H^{\rm L(1)}_{\mu\nu}[G h^{(1)}] + H^{\rm NL(1)}_{\mu\nu}[G h^{(1)}] \right) = 0,
\end{equation}
where $G^{(2)}_{\mu\nu}$ denotes the quadratic part of the Einstein tensor in the metric perturbation, given by~\cite{Wald:1984rg}
\begin{align}
	G^{(2)}_{\mu\nu} [h]
    =&\frac{1}{2}h^{\alpha\beta}\bar{\nabla}_{\mu}\bar{\nabla}_{\nu}h_{\alpha\beta}+\frac{1}{4}\bar{\nabla}_{\mu}h_{\alpha\beta}\bar{\nabla}_{\nu}h^{\alpha\beta} + \bar{\nabla}^{\alpha}h^{\beta}_{\nu}\bar{\nabla}_{[\alpha}h_{\beta]\mu} - h^{\alpha\beta}\bar{\nabla}_{\alpha}\bar{\nabla}_{(\mu}h_{\nu)\beta} \nonumber\\ &+\frac{1}{2}\bar{\nabla}_{\alpha}(h^{\alpha\beta}\bar{\nabla}_{\beta}h_{\mu\nu})-\frac{1}{4}\bar{\nabla}_{\alpha}h_{\mu\nu}\bar{\nabla}^{\alpha}h^{\beta}_{\beta}-\left( \bar{\nabla}_{\beta}h^{\alpha\beta} - \frac{1}{2}\bar{\nabla}^{\alpha}h_{\beta}^{\beta} \right)\bar{\nabla}_{(\mu}h_{\nu)\alpha}\nonumber\\
	&-\eta_{\mu\nu}\bigg(
    \frac{1}{2}h^{\alpha\beta}\bar{\square}h_{\alpha\beta} + \frac{1}{2}h^{\alpha\beta}\bar{\nabla}_{\alpha}\bar{\nabla}_{\beta}h^{\gamma}_{\gamma}-h^{\alpha\beta}\bar{\nabla}_{\alpha}\bar{\nabla}^{\gamma}h_{\beta\gamma} + \frac{3}{8}\bar{\nabla}_{\alpha}h_{\beta\gamma}\bar{\nabla}^{\alpha}h^{\beta\gamma} \nonumber\\ &-\frac{1}{4}\bar{\nabla}^{\alpha}h^{\beta\gamma}\bar{\nabla}_{\beta}h_{\alpha\gamma} + \frac{1}{2}\bar{\nabla}_{\alpha}h^{\alpha\beta}\bar{\nabla}_{\beta}h^{\gamma}_{\gamma}-\frac{1}{8}\bar{\nabla}_{\alpha}h_{\beta}^{\beta}\bar{\nabla}^{\alpha}h^{\gamma}_{\gamma}- \frac{1}{2}\bar{\nabla}_{\beta}h^{\alpha\beta}\bar{\nabla}^{\gamma}h_{\alpha\gamma} \bigg)
\end{align}
with $h^{\mu\nu}=\eta^{\mu\alpha}\eta^{\nu\beta}h_{\alpha\beta}$.
In Eq.~\eqref{eq:eom2}, the linear parts of $H^{\rm L}_{\mu\nu}$ and $H^{\rm NL}_{\mu\nu}$ in the metric perturbation read
\begin{align}
    \label{eq:HL1}
	H^{{\rm L}(1)}_{\mu\nu}[h]
    =& 2\bar{c}_1 \left(\eta_{\mu\nu}\bar{\square} \mathcal{R}^{(1)}[h] - \bar{\nabla}_\mu\bar{\nabla}_\nu \mathcal{R}^{(1)}[h] \right) \nonumber\\
	&+\bar{c}_2 \left( \bar{\square} \mathcal{R}_{\mu\nu}^{(1)}[h]  +
    \frac{1}{2}\eta_{\mu\nu}\bar{\square} \mathcal{R}^{(1)}[h] - \bar{\nabla}^\alpha\bar{\nabla}_\mu \mathcal{R}^{(1)}_{\alpha\nu}[h] - \bar{\nabla}^\alpha\bar{\nabla}_\nu \mathcal{R}_{\alpha\mu}^{(1)}[h] \right)
\end{align}
and
\begin{align}
	\label{eq:HNL1}
	H^{{\rm NL}(1)}_{\mu\nu}[h]
    =& -2\alpha \left(\eta_{\mu\nu}\bar{\square}
    - \bar{\nabla}_\mu\bar{\nabla}_\nu \right)\ln\left(-\frac{\bar{\square}}{\mu^2}\right)\mathcal{R}^{(1)}[h]\nonumber\\
	&-\beta \left( \delta^\alpha_\mu \delta_{\nu}^{\beta}\bar{\square} +
    \eta_{\mu\nu}\bar{\nabla}^\alpha\bar{\nabla}^\beta - \delta^\alpha_\mu\bar{\nabla}^\beta\bar{\nabla}_\nu - \delta^\alpha_\nu\bar{\nabla}^\beta\bar{\nabla}_\mu \right)\ln\left(-\frac{\bar{\square}}{\mu^2}\right)\mathcal{R}^{(1)}_{\alpha\beta}[h]\nonumber\\
    &-2\gamma \left( \delta^\alpha_\mu \eta_{\nu\sigma} + \delta^\alpha_\nu \eta_{\mu\sigma}\right) \bar{\nabla}^\beta \bar{\nabla}_\tau \ln\left(-\frac{\bar{\square}}{\mu^2}\right)\mathcal{R}^{\sigma\tau(1)}_{\alpha\beta}[h].
\end{align}

Let us now solve Eqs.~\eqref{eq:eom1} and \eqref{eq:eom2} to obtain $h^{(1)}_{\mu\nu}$ and $h^{(2)}_{\mu\nu}$, respectively.
In Eq.~\eqref{eq:eom1}, $h^{(1)}_{\mu\nu}$ is found to be
\begin{equation}
	\label{eq:h1}
	h_{tt}^{(1)} = h_{rr}^{(1)} = \frac{2M}{r},
\end{equation}
where $H^{\rm L(0)}_{\mu\nu}[\eta]=H^{\rm NL(0)}_{\mu\nu}[\eta]=0$.
Next,
in Eq.~\eqref{eq:eom2}, the nontrivial quantum contributions come from the third term in Eq.~\eqref{eq:HNL1} because the other terms are  $\mathcal{O}(G^3)$ which is beyond our approximation.
In particular, the nontrivial components of the Riemann-squared term $\mathcal{R}^{\sigma\tau(1)}_{\alpha\beta}[Gh^{(1)}]$ are proportional to $r^{-3}$.
By the use of the calculation~\cite{Calmet:2019eof}
\begin{align}
    \label{eq:ln}
    \ln\left(-\frac{\bar{\square}}{\mu^2}\right) \frac{1}{r^3}
    =& \lim_{\epsilon\to 0^+} \left\{ \frac{1}{r}\int_{\sqrt{\epsilon r}}^{\infty}\frac{\dd r'}{r'^2(r+r')} - \frac{1}{r}\int_{\sqrt{\epsilon r}}^{r-\sqrt{\epsilon r}}\frac{\dd r'}{r'^2(r-r')} - \frac{1}{r}\int_{r+\sqrt{\epsilon r}}^{\infty}\frac{\dd r'}{r'^2(r'-r)} \right. \nonumber\\
    &\left. - \frac{2}{r^3}\left[ \gamma_{\rm E} + \ln (\mu\epsilon) \right] \right\} \nonumber\\
    =& - \frac{2}{r^3} \left[ \ln (\mu r) + \gamma_{\rm E} -1 \right],
\end{align}
where $\sqrt{\epsilon r}$ is a regulator~\cite{Delgado:2022pcc} and $\gamma_{\rm E}$ is the Euler-Mascheroni constant, we obtain  $h^{(2)}_{\mu\nu}$ as
\begin{equation}
    \label{eq:h2}
    h^{(2)}_{tt} = \frac{512\pi\gamma M}{3r^3}, \quad
    h^{(2)}_{rr} = \frac{4M^2}{r^2} + \frac{256\pi\gamma M}{r^3}.
\end{equation}
As a result, combining Eqs.~\eqref{eq:h1} and \eqref{eq:h2}, we find the quantum-corrected metric~\eqref{eq:metric perturb} as
\begin{align}
    g_{tt} =& -1 + \frac{2GM}{r} + \frac{512\pi\gamma G^2M}{3r^3} + \mathcal{O}(G^3), \label{eq:gtt}\\
    g_{rr} =& 1 + \frac{2GM}{r} + \frac{4G^2M^2}{r^2} + \frac{256\pi\gamma G^2M}{r^3} + \mathcal{O}(G^3). \label{eq:grr}
\end{align}
Consequently, in Eqs.~\eqref{eq:gtt} and \eqref{eq:grr}, the quantum corrections appear at order $r^{-3}$ and they are controlled by the parameter $\gamma$ which is determined by the number of massless quantum fields in Eq.~\eqref{eq:gamma}; in this context, $\gamma$ acts as \textit{quantum hair}.

The present quantum-corrected geometry possesses an inner horizon depending on the value of quantum hair $\gamma$.
For negative $\gamma$, we find two black-hole horizons as well as two Killing horizons.
Explicitly, solving $g^{rr}=0$, we obtain the inner and outer black-hole horizons at
\begin{align}
    r_{-} &= \frac{2GM}{3} + \frac{4GM}{3} \cos \left[ \frac{1}{3}\arccos\left( 1+\frac{432\pi\gamma}{GM^2} \right) \right], \label{eq:r-}\\
    r_+   &= \frac{2GM}{3} - \frac{4GM}{3} \cos \left[\frac{1}{3}\left( \pi+ \arccos\left( 1+\frac{432\pi\gamma}{GM^2} \right) \right) \right] \label{eq:r+}
\end{align}
for $-\frac{GM^2}{216\pi}<\gamma<0$, respectively.
Requiring $g_{tt}=0$ yields the inner and outer Killing horizons at
\begin{align}
    r_{k-} &= \frac{2GM}{3} - \frac{4GM}{3} \cos \left[\frac{1}{3}\left( \pi+ \arccos\left( 1+\frac{288\pi\gamma}{GM^2} \right) \right) \right], \label{eq:rk-}\\
    r_{k+} &= \frac{2GM}{3} + \frac{4GM}{3} \cos \left[ \frac{1}{3}\arccos\left( 1+\frac{288\pi\gamma}{GM^2} \right) \right] \label{eq:rk+}
\end{align}
for $-\frac{GM^2}{144\pi}<\gamma<0$, respectively.
Note that the inner Killing horizon emerges inside the inner horizon, whereas the outer Killing horizon emerges outside the outer horizon, $\textit{i.e.},$ $r_{k-}<r_{-}<r_+<r_{k+}$.
Unexpectedly, the Kretschmann scalar diverges at the two Killing horizons, indicating the presence of a naked curvature singularity at the outer Killing horizon.
On the other hand, for $\gamma>0$, there exist the single black-hole horizon~\eqref{eq:r+} and single Killing horizon~\eqref{eq:rk+}, satisfying $r_{k+} < r_+$.
Despite the divergent curvature scalar at the single Killing horizon, it is fortunately inside the black-hole horizon.
To observe the weak cosmic censorship~\cite{Penrose:1964wq,Penrose:1969pc} and strong cosmic censorship~\cite{Poisson:1989zz,Poisson:1990eh,Ori:1991zz,Bonanno:2025bgc}, we will restrict our analysis to $\gamma>0$ in our calculations.

\section{Deflection angles in the strong-field limit}
\label{sec:deflections}
\begin{figure}[t]
\centering
\includegraphics[width=0.8\textwidth]{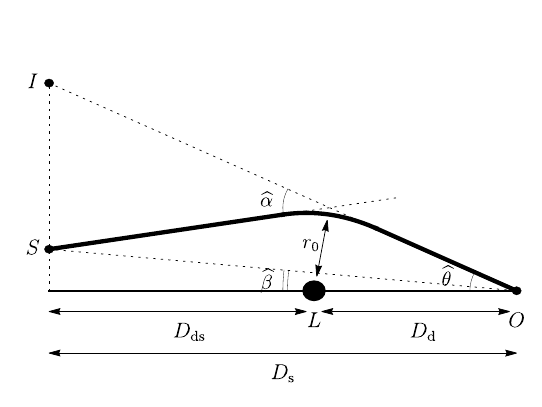}
\caption{The observer $O$, lens $L$, source $S$, and image $I$ are displayed in the lens diagram.
The source and image are at angular positions $\widehat{\beta}$ and $\widehat{\theta}$, respectively.
The light ray emitted from $S$ is bent by the deflection angle $\widehat{\alpha}$ along with the closest approach distance $r_0$.
The distance between the lens and the source-image plane is denoted by $D_{\rm ds}$, while the distance from the observer to the plane is given by $D_{\rm s}$. The distance between the observer and the lens is represented by $D_{\rm d}$.
}
\label{fig:schematics}
\end{figure}

To investigate the quantum-corrected deflection angle, we consider static and spherically symmetric spacetime described as
\begin{equation}
    \label{eq:metric}
    \dd s^2 = -A(r)\dd t^2 + B(r)\dd r^2 + C(r)\dd\Omega^2,
\end{equation}
where the metric functions are given by $A(r) = -g_{tt}(r)$, $B(r) = g_{rr}(r)$ in Eqs.~\eqref{eq:gtt} and \eqref{eq:grr}, and $C(r)=r^2$.
In the metric~\eqref{eq:metric}, a null particle moving on the equatorial plane, \textit{i.e.}, $\theta=\frac{\pi}{2}$, obeys the geodesic equation $u^\mu\nabla_\mu u^\nu =0$, where $u^\mu$ denotes the four-velocity of the null particle.
From the geodesic equation, we can define two conserved quantities: the particle's energy $E$ and angular momentum $L$ as
\begin{equation}
    \label{eq:EnL}
    E=A(r)\dot{t}, \quad L=C(r)\dot{\phi},
\end{equation}
where the overdot denotes a derivative with respect to the affine parameter.
Imposing the null condition $u^\mu u_\mu =0$ then yields
\begin{equation}
    \label{eq:radial_eq}
    \left( \dv{r}{\phi} \right)^2 = \frac{C(r)}{B(r)} \left[ \frac{C(r)}{b^2 A(r)} - 1 \right]
\end{equation}
with the impact parameter defined as $b =\frac{L}{E}$.
Note that the expression on the right-hand side in Eq.~\eqref{eq:radial_eq} plays the role of an effective potential~\cite{Misner:1973prb}.
The schematic diagram of geodesic orbit governed by this radial equation~\eqref{eq:radial_eq} is illustrated in Fig.~\ref{fig:schematics}.
The distance of the closest approach from the lens is denoted by $r_0$; at $r=r_0$, one has $\dv{r}{\phi}=0$, implying that
\begin{equation}
    \label{eq:b}
    b^2=\frac{C(r_0)}{A(r_0)}
\end{equation}
from Eq.~\eqref{eq:radial_eq}.
The condition of the unstable extremum for the effective potential in Eq.~\eqref{eq:radial_eq} is given by~\cite{Bozza:2002zj,Claudel:2000yi}
\begin{equation}
    \label{eq:rps_condition}
    \frac{C'(r_{\rm ps})}{C(r_{\rm ps})}=\frac{A'(r_{\rm ps})}{A(r_{\rm ps})},
\end{equation}
where $r_{\rm ps}$ characterizes the photon sphere.
As $r_0$ decreases, the light ray in Fig.~\ref{fig:schematics} would be bent more, and subsequently when $r_0=r_{\rm ps}$, the ray enters a circular orbit around the lens, which cannot reach to the observer.

The deflection angle $\widehat{\alpha}$ which quantifies the bending of a light ray emitted from the source and passing near the lens before reaching the observer can be expressed as a function of the closest approach distance $r_0$ as
\begin{equation}
    \label{eq:alphahat}
    \widehat{\alpha}(r_0) = I(r_0) -\pi,
\end{equation}
where $I(r_0)$ denotes the azimuthal shift of the light ray, which is given by~\cite{Weinberg:1972kfs}
\begin{equation}
    \label{eq:I}
    I(r_0) = 2\int^\infty_{r_0} \dv{\phi}{r} \dd r = \int^\infty_{r_0} \frac{2\sqrt{B(r_0)}\dd r}{\sqrt{C(r)}\sqrt{ \frac{C(r)A(r_0)}{C(r_0)A(r)} -1}}.
\end{equation}
It is straightforward to see that the integral~\eqref{eq:I} increases as $r_0$ decreases and diverges in the limit $r_0\to r_{\rm ps}$.
In the weak-field limit ($r_0 \gg GM$), Eq.~\eqref{eq:alphahat} can be expanded as
\begin{equation}
    \label{eq:alpha_weak}
    \widehat{\alpha}_{\rm w} (r_0) = \frac{4GM}{r_0} + \frac{4G^2M^2}{r_0^2}\left(\frac{15\pi}{16}-1\right) + \frac{512\pi\gamma G^2M}{r_0^3},
\end{equation}
which is calculated up to $\mathcal{O}(G^2)$.
In Eq.~\eqref{eq:alpha_weak}, the first two terms represent the classical contributions while the last term encodes the quantum corrections proportional to $\gamma$.

In order to describe the deflection angle~\eqref{eq:alphahat} in the strong-field limit ($r_0\gtrsim GM$), we employ the method proposed by Bozza~\cite{Bozza:2002zj}, where Eq.~\eqref{eq:alphahat} is expanded around $r_{\rm ps}$.
Firstly, we consider a change of variable from $r$ to $z=1-\frac{r_0}{r}$ in Eq.~\eqref{eq:I} so that Eq.~\eqref{eq:I} becomes
\begin{equation}
    \label{eq:Iz}
    I(r_0) = \int^1_0 \dd z \frac{R(z,r_0)}{\sqrt{f(z,r_0)}},
\end{equation}
where
\begin{equation}
    \label{eq:Rf}
    R(z,r_0) = \frac{2r_0\sqrt{A(z)B(z)C(0)}}{(1-z)^2C(z)}, \quad
    f(z,r_0) = A(0) - \frac{A(z)C(0)}{C(z)}.
\end{equation}
Here, $A(0)=A(z=0)=1-\frac{2GM}{r_0}-\frac{512\pi\gamma G^2M}{3r_0^3}$ and $C(0)=C(z=0)=r_0^2$.
In Eq.~\eqref{eq:Iz}, the integrand diverges at $z=0$ due to $f(z,r_0)$ which can be expanded near $z=0$ as
\begin{equation}
    \label{}
    f(z,r_0) = f_1(r_0)z + f_2(r_0)z^2 + \mathcal{O}(z^3),
\end{equation}
where
\begin{align}
    \label{}
    f_1(r_0) =& 2 - \frac{6GM}{r_0} - \frac{2560\pi\gamma G^2M}{3r_0^3},\\
    f_2(r_0) =& -1 + \frac{6GM}{r_0} + \frac{5120\pi\gamma G^2M}{3r_0^3}.
\end{align}
Let us now introduce a new function $f_{\rm D}(z,r_0)\equiv f_1(r_0)z + f_2(r_0)z^2$.
Then, we can rewrite Eq.~\eqref{eq:Iz} as
\begin{equation}
    \label{eq:ItoIRnID}
    I(r_0) = I_{\rm R}(r_0) + I_{\rm D}(r_0),
\end{equation}
where
\begin{equation}
    \label{eq:IRnID}
    I_{\rm R}(r_0) = \int^1_0 \dd z \left[ \frac{R(z,r_0)}{\sqrt{f(z,r_0)}} - \frac{R(0,r_{\rm ps})}{\sqrt{f_{\rm D}(z,r_0)}} \right],\quad
    I_{\rm D}(r_0) = \int^1_0 \dd z \frac{R(0,r_{\rm ps})}{\sqrt{f_{\rm D}(z,r_0)}},
\end{equation}
and $R(0,r_{\rm ps})=2r_{\rm ps}\sqrt{\frac{A(r_{\rm ps})B(r_{\rm ps})}{C(r_{\rm ps})}}$.
Note that $I_{\rm R}(r_0)$ in Eq.~\eqref{eq:IRnID} remains finite as $r_0\to r_{\rm ps}$ and it can be expanded as~\cite{Iyer:2006cn,Tsukamoto:2016qro,Tsukamoto:2016jzh}
\begin{equation}
    \label{eq:IRexpand}
    I_{\rm R}(r_0) = I_{\rm R}(r_{\rm ps}) + \mathcal{O}\left(\left(\frac{r_0}{r_{\rm ps}}-1\right) \ln \left(\frac{r_0}{r_{\rm ps}}-1\right)\right),
\end{equation}
which has to be evaluated numerically.
On the other hand, $I_{\rm D}(r_0)$ in Eq.~\eqref{eq:IRnID} diverges as $r_0\to r_{\rm ps}$ but it can be integrated analytically in $z$ as
\begin{equation}
    \label{eq:IDraw}
    I_{\rm D}(r_0) = \frac{2R(0,r_{\rm ps})}{\sqrt{f_2(r_0)}} \ln \left[ \frac{\sqrt{f_2(r_0)}+\sqrt{f_1(r_0)+f_2(r_0)}}{\sqrt{f_1(r_0)}} \right],
\end{equation}
which also can be expanded as~\cite{Iyer:2006cn,Tsukamoto:2016qro,Tsukamoto:2016jzh}
\begin{equation}
    \label{eq:IDexpand}
    I_{\rm D}(r_0) = - \frac{R(0,r_{\rm ps})}{\sqrt{f_2(r_{\rm ps})}} \left[ \ln\left( \frac{r_0}{r_{\rm ps}} -1 \right) - \ln 2 \right] + \mathcal{O}\left(\left(\frac{r_0}{r_{\rm ps}}-1\right) \ln \left(\frac{r_0}{r_{\rm ps}}-1\right)\right).
\end{equation}
Plugging Eqs.~\eqref{eq:IRexpand} and \eqref{eq:IDexpand} into Eqs.~\eqref{eq:alphahat} and \eqref{eq:ItoIRnID}, the deflection angle in the strong-field limit takes the form~\cite{Liu:2024wal}
\begin{align}
    \widehat{\alpha}_{\rm s}(r_0) &\simeq - \frac{R(0,r_{\rm ps})}{\sqrt{f_2(r_{\rm ps})}} \left[ \ln\left( \frac{r_0}{r_{\rm ps}} -1 \right) - \ln 2 \right] + I_{\rm R}(r_{\rm ps}) - \pi \nonumber\\
    &\equiv - a_{\rm s} \ln\left( \frac{r_0}{r_{\rm ps}} -1 \right) + b_{\rm s}, \label{eq:alpha_strong}
\end{align}
where $a_{\rm s}$ and $b_{\rm s}$ are the strong-field coefficients depending on $\gamma$.

\begin{figure}[t]
	\centering
	\subfigure[~The radius of photon sphere]{\includegraphics[width=0.48\textwidth]{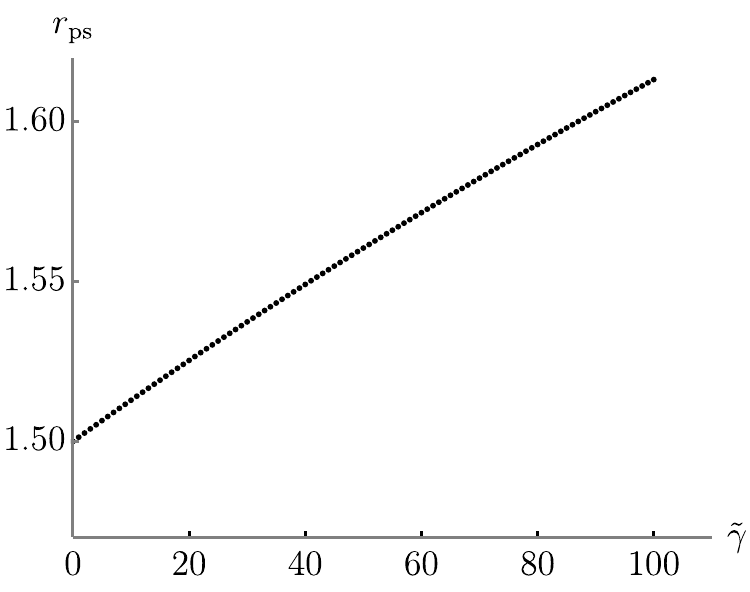}\label{fig:rps}}\qquad
	\subfigure[~The deflection angle in the strong field limit]{\includegraphics[width=0.47\textwidth]{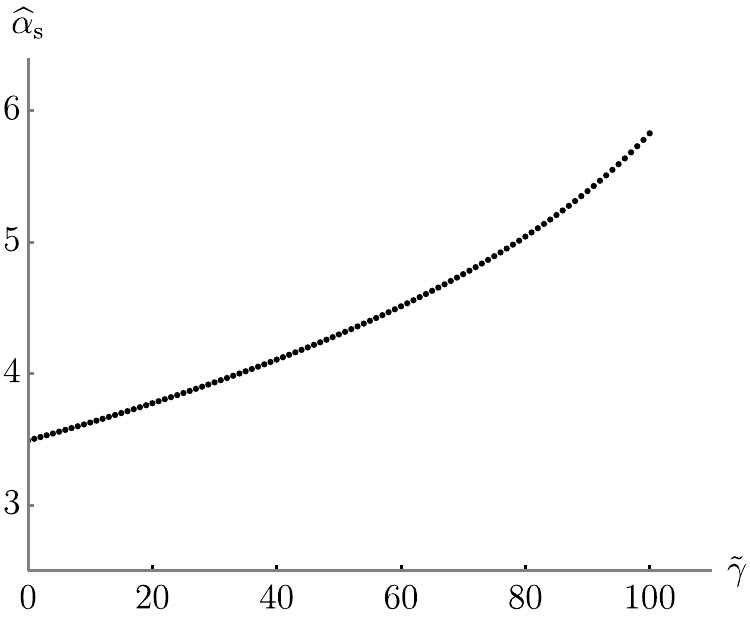}\label{fig:alpha}}
	\caption{
        The radius of photon sphere $r_{\rm ps}$ and the deflection angle in the strong field limit $\widehat{\alpha}_{\rm s}$ are displayed in Figs.~\ref{fig:rps} and \ref{fig:alpha}, respectively, as functions of the rescaled nonlocal Wilson coefficient $\tilde{\gamma} \equiv 11520\pi^2 \gamma$.
        In Fig.~\ref{fig:rps}, $r_{\rm ps}$ increases monotonically with $\tilde{\gamma}$, passing through $r_{\rm ps}=1.5$ for $\tilde{\gamma}=0$.
        In Fig.~\ref{fig:alpha}, $\widehat{\alpha}_{\rm s}$ is evaluated at the fixed radius of closest approach as $r_0\approx 1.64325$, where for $\tilde{\gamma}=0$, $\widehat{\alpha}_{\rm s}\approx 3.49092$. In these calculations, we set $G=0.5$ and $M=1$ for simplicity.
            }
\end{figure}
We now examine the dependence of $\gamma$ on the deflection angle in Eq.~\eqref{eq:alpha_strong}.
For the quantum-corrected metric functions~\eqref{eq:gtt} and \eqref{eq:grr}, we numerically evaluate $r_{\rm ps}$ in Eq.~\eqref{eq:rps_condition} and Eq.~\eqref{eq:alpha_strong}, as illustrated in Figs.~\ref{fig:rps} and \ref{fig:alpha}, respectively.
In Figs.~\ref{fig:rps} and \ref{fig:alpha}, we redefine $\gamma$ as $\tilde{\gamma}\equiv 11520\pi^2\gamma$ varying over the range $\tilde{\gamma}\in [0,100]$.
As shown in Fig.~\ref{fig:rps}, $r_{\rm ps}$ increases monotonically as $\tilde{\gamma}$ increases.
Similarly, in Fig.~\ref{fig:alpha}, the deflection angle $\widehat{\alpha}_{\rm s}$ also increases with $\tilde{\gamma}$, while the strong-field coefficients $a_{\rm s}$ and $b_{\rm s}$ in Eq.~\eqref{eq:alpha_strong} slightly decrease as $\tilde{\gamma}$ increases.
Note that as $\tilde{\gamma}$ varies from 0 to 100, $a_{\rm s}$ decreases gradually from 1.67774 to 1.58958, while $b_{\rm s}$ decreases slightly from -0.44952 to -0.50695.
Despite these small decreases in the coefficient $a_{\rm s}$ and $b_{\rm s}$, $\widehat{\alpha}_{\rm s}$ as well as $r_{\rm ps}$ manifestly increase depending on quantum hair.

\section{Lensing observables in the strong-field limit}
\label{sec:observables}
In this section, we investigate quantum corrections to lensing observables in the regime of strong gravitational lensing.
In Fig.~\ref{fig:schematics}, the lens equation is given by~\cite{Virbhadra:1999nm}
\begin{equation}
    \label{eq:lens_eq}
    \tan\widehat{\beta} = \tan\widehat{\theta} - \frac{D_{\rm ds}}{D_{\rm s}} \left[ \tan\widehat{\theta} + \tan (\widehat{\alpha}_{\rm s}-\widehat{\theta}) \right],
\end{equation}
where $\widehat{\beta}$ and $\widehat{\theta}$ denote the angular positions of the source and the image, respectively, and $D_{\rm ds}$ and $D_{\rm s}$ are the lens-to-source and observer-to-source distances.
In Eq.~\eqref{eq:lens_eq}, both $\widehat{\alpha}_{\rm s}$ in Eq.~\eqref{eq:alpha_strong} and $\widehat{\theta}$ depend on $r_0$.
In particular, $\widehat{\theta}$ is related to $r_0$ via
\begin{equation}
    \label{eq:theta}
    \widehat{\theta} = \arcsin \left[ \frac{b}{D_{\rm d}} \right] = \arcsin \left[ \frac{\sqrt{C(r_0)}}{D_{\rm d}\sqrt{A(r_0)}} \right],
\end{equation}
where $b$ is replaced by Eq.~\eqref{eq:b} and $D_{\rm d}$ denotes the distance between the lens and the observer.
Let us now consider the situation in which the light ray winds exactly $n$ times around the lens before reaching the observer, corresponding to $r_0=r_0^{(n)}$.
In this situation, the strong-field deflection angle~\eqref{eq:alpha_strong} becomes $\widehat{\alpha}_{\rm s}\left(r_0^{(n)}\right)=2n\pi$ for
$r_0^{(n)} = r_{\rm ps} \left[ 1+\exp\left(\frac{b_{\rm s}-2n\pi}{a_{\rm s}}\right) \right]$
and the corresponding $\widehat{\theta}_0^{(n)}$, defined by plugging $r_0^{(n)}$ into Eq.~\eqref{eq:theta}, is always equal to $\widehat{\beta}$ in accordance with Eq.~\eqref{eq:lens_eq}.

For $r_0=r_0^{(n)}+\delta$ that slightly deviates from $r_0^{(n)}$ where $\delta \ll r_0^{(n)}$, the deflection angle~\eqref{eq:alpha_strong} can be expanded as
\begin{equation}
    \label{eq:alpha_strong_expand}
    \widehat{\alpha}_{\rm s} (r_0) = 2n\pi + \frac{a_{\rm s}}{r_{\rm ps}} \exp\left[ -\frac{b_{\rm s}-2n\pi}{a_{\rm s}} \right] \delta + \mathcal{O}(\delta^2).
\end{equation}
Then, the corresponding image position~\eqref{eq:theta} becomes
\begin{equation}
    \label{eq:theta_expand}
    \widehat{\theta}^{(n)}
    = \widehat{\theta}_0^{(n)}
    + \frac{\sqrt{C\left(r_0^{(n)}\right)}}{2\sqrt{D_{\rm d}^2A\left(r_0^{(n)}\right)-C\left(r_0^{(n)}\right)}} \left[ \frac{C'\left(r_0^{(n)}\right)}{C\left(r_0^{(n)}\right)} - \frac{A'\left(r_0^{(n)}\right)}{A\left(r_0^{(n)}\right)} \right]\delta + \mathcal{O}(\delta^2),
\end{equation}
whose asymptotic limit as $n\to\infty$ is given by $\widehat{\theta}^{(\infty)} = \arcsin \left[ \frac{\sqrt{C(r_{\rm ps})}}{D_{\rm d}\sqrt{A(r_{\rm ps})}} \right]$.
For $n=1$ and $n\to\infty$ in Eq.~\eqref{eq:theta_expand}, the angular separation between the outermost ($n=1$) and the asymptotic ($n\to\infty$) images is defined as~\cite{Bozza:2002zj}
\begin{equation}
    \label{eq:s}
    s = \widehat{\theta}^{(1)} - \widehat{\theta}^{(\infty)}.
\end{equation}
The relative magnification of the outermost image is defined as~\cite{Virbhadra:1999nm,Bozza:2002zj}
\begin{equation}
    \label{eq:mu_rel}
    \mu_{\rm rel} = \frac{\mu^{(1)}}{\sum_{n=2}^{\infty}\mu^{(n)}}
\end{equation}
which characterizes the magnification ratio between the outermost image and all the other relativistic images ($n\ge 2$).
In Eq.~\eqref{eq:mu_rel}, the magnification of each individual image $\mu^{(n)}$ is
\begin{equation}
    \label{eq:mu_n}
    \mu^{(n)} = \left( \frac{\sin\widehat{\beta}}{\sin\widehat{\theta}^{(n)}} \dv{\widehat{\beta}}{\widehat{\theta}^{(n)}} \right)^{-1},
\end{equation}
where $\widehat{\beta}$ is determined by Eq.~\eqref{eq:lens_eq}.

\begin{figure}[t]
	\centering
	\subfigure[~The angular separation]{\includegraphics[width=0.49\textwidth]{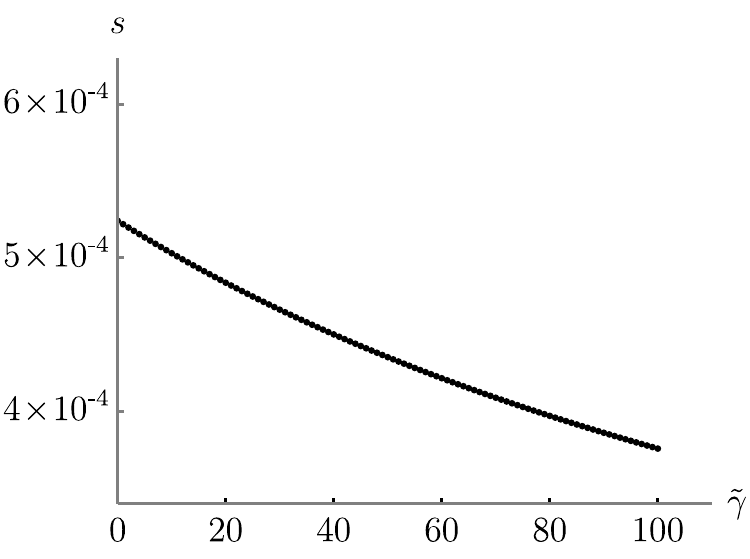}\label{fig:s}}\qquad
	\subfigure[~The relative magnification]{\includegraphics[width=0.46\textwidth]{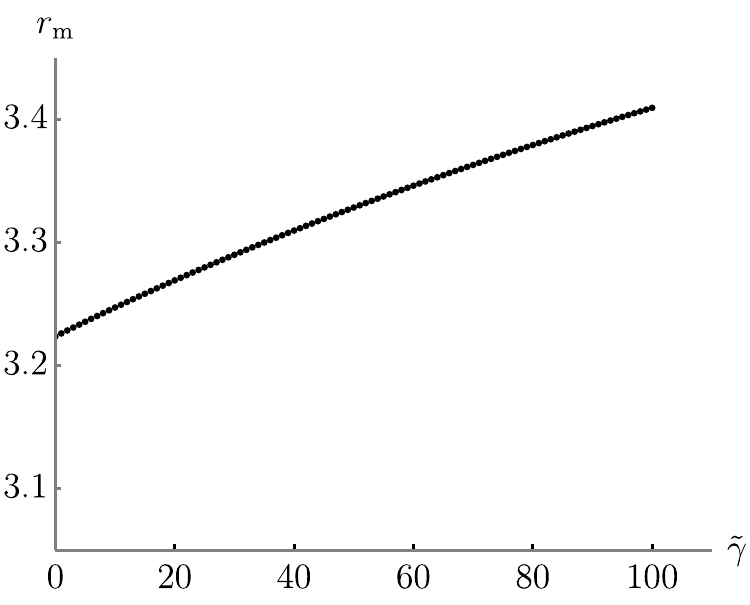}\label{fig:rm}}
	\caption{
        The angular separation $s$ and the relative magnification $\mu_{\rm rel}$ are shown in Figs.~\ref{fig:s} and \ref{fig:rm}, respectively, as functions of the rescaled nonlocal Wilson coefficient $\tilde{\gamma} \equiv 11520\pi^2 \gamma$.
        In Fig.~\ref{fig:rm}, we use the rescaled relative magnification as $r_{\rm m}\equiv2.5 \log_{10}\left(\mu_{\rm rel}\right)$.
		In these calculations, we set $G=0.5$, $M=1$, $\delta=10^{-2}$ and $\widehat{\beta}=0.5$ for simplicity.
            }
\end{figure}\label{fig:strong}
In Figs.~\ref{fig:s} and \ref{fig:rm}, the angular separation~\eqref{eq:s} and the rescaled relative magnification defined by Eq.~\eqref{eq:mu_rel} are plotted with respect to $\tilde{\gamma}$.
In Fig.~\ref{fig:s}, the angular separation $s$ monotonically decreases with increasing $\tilde{\gamma}$, indicating that all relativistic images between the outermost and asymptotic images become increasingly concentrated in a narrower angular region as $\tilde{\gamma}$ increases.
In contrast, in Fig.~\ref{fig:rm}, the rescaled relative magnification $r_{\rm m}$ increases with $\tilde{\gamma}$, implying that the brightness contrast between the outermost image and the other relativistic images becomes more pronounced as $\tilde{\gamma}$ grows.
Consequently, the two strong-field lensing observables exhibit distinct behaviors depending on quantum hair.

\begin{table}[t]
\centering
\begin{tabular}{c|c c c c}
  \hline
  $~\gamma_{\rm re}~$ & $~~r_{\rm ps}/R_{\rm S}~~$  & $~\widehat{\theta}^{(\infty)}~(\mu \text{arcsecs})~$ & $~s~(\mu \text{arcsecs})~$ & $~r_{\rm m}~(\text{magnitudes})~$ \\ [0.5ex]
  \hline
  0      & 1.5     & 26.5478 & 0.013017    & 8.13222 \\
  10     & 2.22229 & 30.6190 & 0.0021195   & 9.94795 \\
  20     & 2.57552 & 33.6482 & 0.0014995   & 10.3595 \\
  50     & 3.22009 & 39.6272 & 0.0010037   & 10.8219 \\
  100    & 3.87527 & 45.9717 & 0.000818491 & 11.0999 \\
  \hline
\end{tabular}
\caption{The radius of photon sphere and strong-field lensing observables for the parameters $M=4.4\times 10^6 M_\odot$ and $D_{\rm d}=8.5$kpc with respect to $\gamma_{\rm re}=\frac{\hbar G\gamma}{c^5}$, where $R_{\rm S}=\frac{2GM}{c^2}$ denotes the Schwarzschild radius.}
\label{tab:strong}
\end{table}
To explore the observational dependence of quantum hair in an astrophysical situation, let us consider the black hole at the center of our galaxy as the gravitational lens, whose mass is $M=4.4\times 10^6M_\odot$ and distance from the observer is approximately $D_{\rm d}=8.5$kpc as considered in Ref.~\cite{Soares:2025hpy}.
In this case, we can approximate $D_{\rm d}\tan\widehat{\theta}\simeq D_{\rm d}\widehat{\theta}$, simplifying the lens equation~\eqref{eq:lens_eq} as
\begin{equation}\label{eq:lens_eq_app}
    \widehat{\beta} = \widehat{\theta} - \frac{D_{\rm ds}}{D_{\rm s}} \Delta \widehat{\alpha}_{\rm s}^{(n)},
\end{equation}
where $\Delta \widehat{\alpha}_{\rm s}^{(n)} = \widehat{\alpha}_{\rm s}-2n\pi$.
As found in Ref.~\cite{Bozza:2002zj}, the strong-field lensing observables in Eqs.~\eqref{eq:s} and \eqref{eq:mu_rel} can then be approximated by
\begin{equation}\label{eq:strong_obs_app}
    s \simeq \widehat{\theta}^{(\infty)} e^{\frac{\bar{b}_{\rm s}-2\pi}{\bar{a}_{\rm s}}}, \quad \mu_{\rm rel} \simeq e^{\frac{2\pi}{\bar{a}_{\rm s}}},
\end{equation}
where $\bar{a}_{\rm s}=\frac{a_{\rm s}}{2}$ and $\bar{b}_{\rm s}=\bar{a}_{\rm s}\ln\left(\frac{2 n(r_{\rm ps})}{A(r=r_{\rm ps})}\right)+I_{\rm R}(r_{\rm ps})-\pi$.
To compute Eq.~\eqref{eq:strong_obs_app}, we restore the dimensional units in the metric components.
For example, the $tt$-component in Eq.~\eqref{eq:h1} becomes $Gh^{(1)}_{tt}=\frac{2GM}{c^2}$, while the quantum-corrected term takes the form $G^2h^{(2)}_{tt}=\frac{512\pi\hbar\gamma G^2M}{3c^5r^3}$ in Eq.~\eqref{eq:h2}.
Although the quantum-corrected terms in the metric are extremely small, making their observation challenging in current data such as those reported in Refs.~\cite{EventHorizonTelescope:2022urf,Tiede:2022grp},
the observables in Eq.~\eqref{eq:strong_obs_app} can numerically be evaluated by redefining $\gamma_{\rm re}=\frac{\hbar G\gamma}{c^5}$ as shown in Table.~\ref{tab:strong}.
The classical values of $s$ and $r_{\rm m}$ for $\gamma_{\rm re}=0$ are slightly different from those obtained in Ref.~\cite{Soares:2025hpy} because of the different choice of the background metric.

It is worth noting that, in a similar manner, several lensing observables can also be evaluated in the weak-field limit by using the deflection angle $\widehat{\alpha}_{\rm w}$ in Eq.~\eqref{eq:alpha_weak}.
Along the line of Ref.~\cite{Abe:2010ap}, we consider a bulge star with $D_{\rm d}=D_{\rm ds}=4$kpc and the lens mass $M=0.3M_\odot$.
In this situation, when the source, lens, and observer are perfectly aligned ($\widehat{\beta}=0$), the lens equation~\eqref{eq:lens_eq} reduces to
\begin{equation}\label{eq:lens_eq_app2}
    \widehat{\theta} = \frac{D_{\rm ds}}{D_{\rm s}} \widehat{\alpha}_{\rm w}
\end{equation}
and the corresponding solution $\widehat{\theta}=\widehat{\theta}_{\rm E}$ defines the Einstein ring with its radius $R_{\rm E}\simeq D_{\rm d}\widehat{\theta}_{\rm E}$.
Substituting $\widehat{\theta}\simeq \frac{1}{D_{\rm d}}\sqrt{\frac{C(r_0)}{A(r_0)}}$ in Eq.~\eqref{eq:theta} into Eq.~\eqref{eq:lens_eq_app2}, we obtain the values of $\widehat{\theta}_{\rm E}$ and $R_{\rm E}$, as shown in Table.~\ref{tab:weak}.
Although the classical values ($\gamma_{\rm re}=0$) of $\widehat{\theta}_{\rm E}$ and $R_{\rm E}$ in Table.~\ref{tab:weak} seem to be different from those calculated in Ref.~\cite{Soares:2025hpy}, this discrepancy arises from the different choice of the lens mass.
As a result, the observables increase monotonically with the redefined parameter $\gamma_{\rm re}$ even in the weak-field regime.
\begin{table}[t]
\centering
\begin{tabular}{c|c c}
  \hline
  $~\gamma_{\rm re}~$ & $~~~~\widehat{\theta}_{\rm E}~(\text{arcsecs})~~~~$ & $~~~~R_{\rm E}~(\text{km})~~~~$ \\ [0.5ex]
  \hline
  0      & 5.52605 $\times 10^{-4}$ & 3.30708 $\times 10^8$ \\
  10     & 5.53514 $\times 10^{-4}$ & 3.31253 $\times 10^8$ \\
  20     & 5.54416 $\times 10^{-4}$ & 3.31792 $\times 10^8$ \\
  50     & 5.57079 $\times 10^{-4}$ & 3.33386 $\times 10^8$ \\
  100    & 5.61383 $\times 10^{-4}$ & 3.35961 $\times 10^8$ \\
  \hline
\end{tabular}
\caption{The weak-field lensing observables for the parameters $M=0.3 M_\odot$ and $D_{\rm d}=D_{\rm ds}=4$kpc with respect to $\gamma_{\rm re}=\frac{\hbar G\gamma}{c^5}$.}
\label{tab:weak}
\end{table}

In Tables.~\ref{tab:strong} and \ref{tab:weak}, the lensing observables for such values of $\gamma_{\rm re}$ show significant modifications from their classical results.
For instance, as $\gamma_{\rm re}$ increases from 0 to 100 in Table~\ref{tab:strong}, the photon-sphere radius grows by more than 150\% and the relative magnification increases by about 36\%.
These variations are well within the sensitivity of current observational capabilities, including those of the \textit{Event Horizon Telescope} (EHT) observations of M87*~\cite{EventHorizonTelescope:2019ggy,EventHorizonTelescope:2025vum}.
In Table.~\ref{tab:weak}, the weak-field lensing observables undergo modifications at the level of approximately 1.5\%, which are comparable to the high-precision measurements expected from near-future surveys such as the \textit{Large Synoptic Survey Telescope} (LSST) and \textit{Euclid}~\cite{LSST:2008ijt,Euclid:2024yrr}.

\section{conclusion and discussion}
\label{sec:conclusion}
In this paper, we derived the quantum-corrected black hole metric possessing quantum hair from the effective action developed by Barvinsky and Vilkovisky without invoking either the commutativity of the nonlocal operator with covariant derivatives or the nonlocal Gauss-Bonnet theorem.
The resulting metric is perturbatively obtained on a flat background, which incorporates  quantum corrections that scale as $r^{-3}$.
These quantum corrections to the metric are characterized by the nonlocal Wilson coefficient $\gamma$ due to the nontrivial contributions of Riemann tensor.
Using the quantum-corrected metric, we investigated quantum effects in the strong deflection limit of light, where light rays pass arbitrary close to the photon sphere.
Through numerical computations, we showed that as $\gamma$ increases, the radius of the photon sphere, the deflection angle in the strong-field limit, and the relativistic magnification all increase, whereas the angular separation between the outermost and the asymptotic relativistic images decreases.
These results indicate that the quantum hair $\gamma$ leads to, in principle, distinguishable modifications in strong lensing observables.
Redefining the quantum-hair parameter $\gamma_{\rm re}$, we further evaluated the quantum-corrected observables in the strong-field limit quantitatively.
For large values of $\gamma_{\rm re}$, these observables exhibit significant deviations from the classical results, which are comparable to current observational sensitivities.
Although those for more precise value of quantum hair may be challenging to detect with present facilities, improvements in high-precision lensing measurements from next generation surveys may offer an opportunity to examine such quantum corrections.

Given these results, one might wonder:
how do the photon sphere radius, the strong deflection angle, and the relativistic magnification all increase with increasing $\gamma$, while the angular separation decreases?
This increasing behavior can be understood by the effective mass, defined in Eq.~\eqref{eq:gtt} as $M_{\rm eff}(r)=M\left(1+\frac{512\pi\gamma G^2}{3r^3}\right)$.
The effective mass of the black hole increases linearly with $\gamma$, thereby shifting the radius of the photon sphere outward, which eventually enhances the deflection of light as well as the strong-field deflection angle and the relativistic magnification.
For the angular separation, both angular positions $\widehat{\theta}^{(1)}$ and $\widehat{\theta}^{(\infty)}$ increase with growing $\gamma$; however, $\widehat{\theta}^{(\infty)}$, which is more sensitive to the photon sphere radius than $\widehat{\theta}^{(1)}$, increases more rapidly than the latter.
As a result, the angular separation $s$ exhibits a net decreasing behavior.
Note that these trends with respect to $\gamma$ in the lensing observables are also observed in their approximate expressions in Eq.~\eqref{eq:strong_obs_app}, which are derived under the assumption that the angles in Eqs.~\eqref{eq:lens_eq} and \eqref{eq:theta}, such as $\widehat{\beta}$, $\widehat{\theta}$, and $\Delta \widehat{\alpha}_{\rm s}^{(n)}$ are small~\cite{Bozza:2002zj}.

The reason why $\gamma$ appears as quantum hair in our analysis in contrast to Ref.~\cite{Calmet:2017qqa} is that we do not assume that the nonlocal operator commutes with the covariant derivative in curved spacetime.
This assumption may be questioned because the nonlocal operator can be expressed as an infinite series expansion in powers of $\square$, which does not commute with the covariant derivative in curved backgrounds.
To clarify this issue, let us consider the flat spacetime case as an example.
In flat spacetime, the d'Alembertian operator trivially commutes with ordinary derivative, so one might expect that they would commute as well.
Nevertheless, a direct computation using Eq.~\eqref{eq:ln} reveals that $\left[\ln \left(-\frac{\bar{\square}}{\mu^2} \right), \bar{\nabla}_\mu\right]\frac{1}{r^2}=-\frac{2}{r^3}$, which clearly shows that the nonlocal operator and the ordinary derivative do not commute even in flat spacetime.
For this reason, we avoid assuming the vanishing commutation relation and instead directly vary the Riemann-squared term in the nonlocal effective action, leading to the quantum-corrected black hole metric that carries an explicit dependence on quantum hair.

\acknowledgments
We would like to thank Mungon Nam for exciting discussions.
This research was supported by Basic Science Research Program through the National Research Foundation of Korea(NRF) funded by the Ministry of Education through the Center for Quantum Spacetime (CQUeST) of Sogang University (No. RS-2020-NR049598).
This work was supported by the National Research Foundation of Korea(NRF) grant funded by the Korea government(MSIT).(No. RS-2022-NR069013)


\bibliographystyle{JHEP}       

\bibliography{references}

\providecommand{\href}[2]{#2}\begingroup\raggedright\begin{thebibliography}{10}

\bibitem{Israel:1967wq}
W.~Israel, \emph{{Event horizons in static vacuum space-times}},
  \href{http://dx.doi.org/10.1103/PhysRev.164.1776}{\emph{Phys. Rev.} {\bf 164}
  (1967) 1776--1779}.

\bibitem{Israel:1967za}
W.~Israel, \emph{{Event horizons in static electrovac space-times}},
  \href{http://dx.doi.org/10.1007/BF01645859}{\emph{Commun. Math. Phys.} {\bf
  8} (1968) 245--260}.

\bibitem{Carter:1971zc}
B.~Carter, \emph{{Axisymmetric Black Hole Has Only Two Degrees of Freedom}},
  \href{http://dx.doi.org/10.1103/PhysRevLett.26.331}{\emph{Phys. Rev. Lett.}
  {\bf 26} (1971) 331--333}.

\bibitem{Robinson:1975bv}
D.~C. Robinson, \emph{{Uniqueness of the Kerr black hole}},
  \href{http://dx.doi.org/10.1103/PhysRevLett.34.905}{\emph{Phys. Rev. Lett.}
  {\bf 34} (1975) 905--906}.

\bibitem{Barvinsky:1983vpp}
A.~O. Barvinsky and G.~A. Vilkovisky, \emph{{The Generalized Schwinger-De Witt
  Technique and the Unique Effective Action in Quantum Gravity}},
  \href{http://dx.doi.org/10.1016/0370-2693(83)90506-3}{\emph{Phys. Lett. B}
  {\bf 131} (1983) 313--318}.

\bibitem{Barvinsky:1985an}
A.~O. Barvinsky and G.~A. Vilkovisky, \emph{{The Generalized Schwinger-Dewitt
  Technique in Gauge Theories and Quantum Gravity}},
  \href{http://dx.doi.org/10.1016/0370-1573(85)90148-6}{\emph{Phys. Rept.} {\bf
  119} (1985) 1--74}.

\bibitem{Barvinsky:1987uw}
A.~O. Barvinsky and G.~A. Vilkovisky, \emph{{Beyond the Schwinger-Dewitt
  Technique: Converting Loops Into Trees and In-In Currents}},
  \href{http://dx.doi.org/10.1016/0550-3213(87)90681-X}{\emph{Nucl. Phys. B}
  {\bf 282} (1987) 163--188}.

\bibitem{Barvinsky:1990up}
A.~O. Barvinsky and G.~A. Vilkovisky, \emph{{Covariant perturbation theory. 2:
  Second order in the curvature. General algorithms}},
  \href{http://dx.doi.org/10.1016/0550-3213(90)90047-H}{\emph{Nucl. Phys. B}
  {\bf 333} (1990) 471--511}.

\bibitem{Barvinsky:1990uq}
A.~O. Barvinsky and G.~A. Vilkovisky, \emph{{Covariant perturbation theory. 3:
  Spectral representations of the third order form-factors}},
  \href{http://dx.doi.org/10.1016/0550-3213(90)90048-I}{\emph{Nucl. Phys. B}
  {\bf 333} (1990) 512--524}.

\bibitem{Calmet:2019eof}
X.~Calmet, R.~Casadio and F.~Kuipers, \emph{{Quantum Gravitational Corrections
  to a Star Metric and the Black Hole Limit}},
  \href{http://dx.doi.org/10.1103/PhysRevD.100.086010}{\emph{Phys. Rev. D} {\bf
  100} (2019) 086010}, [\href{http://arxiv.org/abs/1909.13277}{{\tt
  1909.13277}}].

\bibitem{Calmet:2020tlj}
X.~Calmet, R.~Casadio and F.~Kuipers, \emph{{Quantum corrected equations of
  motion in the interior and exterior Schwarzschild spacetime}},
  \href{http://dx.doi.org/10.1103/PhysRevD.102.026018}{\emph{Phys. Rev. D} {\bf
  102} (2020) 026018}, [\href{http://arxiv.org/abs/2007.05416}{{\tt
  2007.05416}}].

\bibitem{Calmet:2021stu}
X.~Calmet, R.~Casadio, S.~D.~H. Hsu and F.~Kuipers, \emph{{Quantum Hair from
  Gravity}},
  \href{http://dx.doi.org/10.1103/PhysRevLett.128.111301}{\emph{Phys. Rev.
  Lett.} {\bf 128} (2022) 111301}, [\href{http://arxiv.org/abs/2110.09386}{{\tt
  2110.09386}}].

\bibitem{Calmet:2023met}
X.~Calmet, R.~Casadio, S.~D.~H. Hsu and F.~Kuipers, \emph{{Quantum hair during
  gravitational collapse}},
  \href{http://dx.doi.org/10.1103/PhysRevD.108.086012}{\emph{Phys. Rev. D} {\bf
  108} (2023) 086012}, [\href{http://arxiv.org/abs/2305.09466}{{\tt
  2305.09466}}].

\bibitem{Cheong:2023oik}
S.~Cheong and W.~Kim, \emph{{Quantum geodesics reflecting the internal
  structure of stars composed of shells}},
  \href{http://dx.doi.org/10.1088/1475-7516/2024/07/089}{\emph{JCAP} {\bf 07}
  (2024) 089}, [\href{http://arxiv.org/abs/2311.13113}{{\tt 2311.13113}}].

\bibitem{Perrucci:2024qrr}
I.~Perrucci, F.~Kuipers and R.~Casadio, \emph{{Quantum gravitational hair in
  gravastars and observational tests}},
  \href{http://dx.doi.org/10.1088/1475-7516/2025/03/005}{\emph{JCAP} {\bf 03}
  (2025) 005}, [\href{http://arxiv.org/abs/2412.04886}{{\tt 2412.04886}}].

\bibitem{Cheong:2025scf}
S.~Cheong, W.~Kim and M.~Nam, \emph{{Einstein ring of dust shells with quantum
  hair}}, \href{http://dx.doi.org/10.1088/1475-7516/2025/07/041}{\emph{JCAP}
  {\bf 07} (2025) 041}, [\href{http://arxiv.org/abs/2504.08343}{{\tt
  2504.08343}}].

\bibitem{Calmet:2017qqa}
X.~Calmet and B.~K. El-Menoufi, \emph{{Quantum Corrections to Schwarzschild
  Black Hole}},
  \href{http://dx.doi.org/10.1140/epjc/s10052-017-4802-0}{\emph{Eur. Phys. J.
  C} {\bf 77} (2017) 243}, [\href{http://arxiv.org/abs/1704.00261}{{\tt
  1704.00261}}].

\bibitem{Calmet:2018elv}
X.~Calmet, \emph{{Vanishing of Quantum Gravitational Corrections to Vacuum
  Solutions of General Relativity at Second Order in Curvature}},
  \href{http://dx.doi.org/10.1016/j.physletb.2018.10.040}{\emph{Phys. Lett. B}
  {\bf 787} (2018) 36--38}, [\href{http://arxiv.org/abs/1810.09719}{{\tt
  1810.09719}}].

\bibitem{Xiao:2021zly}
Y.~Xiao and Y.~Tian, \emph{{Logarithmic correction to black hole entropy from
  the nonlocality of quantum gravity}},
  \href{http://dx.doi.org/10.1103/PhysRevD.105.044013}{\emph{Phys. Rev. D} {\bf
  105} (2022) 044013}, [\href{http://arxiv.org/abs/2104.14902}{{\tt
  2104.14902}}].

\bibitem{Virbhadra:1999nm}
K.~S. Virbhadra and G.~F.~R. Ellis, \emph{{Schwarzschild black hole lensing}},
  \href{http://dx.doi.org/10.1103/PhysRevD.62.084003}{\emph{Phys. Rev. D} {\bf
  62} (2000) 084003}, [\href{http://arxiv.org/abs/astro-ph/9904193}{{\tt
  astro-ph/9904193}}].

\bibitem{Bozza:2002zj}
V.~Bozza, \emph{{Gravitational lensing in the strong field limit}},
  \href{http://dx.doi.org/10.1103/PhysRevD.66.103001}{\emph{Phys. Rev. D} {\bf
  66} (2002) 103001}, [\href{http://arxiv.org/abs/gr-qc/0208075}{{\tt
  gr-qc/0208075}}].

\bibitem{Iyer:2006cn}
S.~V. Iyer and A.~O. Petters, \emph{{Light's bending angle due to black holes:
  From the photon sphere to infinity}},
  \href{http://dx.doi.org/10.1007/s10714-007-0481-8}{\emph{Gen. Rel. Grav.}
  {\bf 39} (2007) 1563--1582}, [\href{http://arxiv.org/abs/gr-qc/0611086}{{\tt
  gr-qc/0611086}}].

\bibitem{Tsukamoto:2016qro}
N.~Tsukamoto, \emph{{Strong deflection limit analysis and gravitational lensing
  of an Ellis wormhole}},
  \href{http://dx.doi.org/10.1103/PhysRevD.94.124001}{\emph{Phys. Rev. D} {\bf
  94} (2016) 124001}, [\href{http://arxiv.org/abs/1607.07022}{{\tt
  1607.07022}}].

\bibitem{Tsukamoto:2016jzh}
N.~Tsukamoto, \emph{{Deflection angle in the strong deflection limit in a
  general asymptotically flat, static, spherically symmetric spacetime}},
  \href{http://dx.doi.org/10.1103/PhysRevD.95.064035}{\emph{Phys. Rev. D} {\bf
  95} (2017) 064035}, [\href{http://arxiv.org/abs/1612.08251}{{\tt
  1612.08251}}].

\bibitem{Soares:2025hpy}
A.~R. Soares, C.~F.~S. Pereira, R.~L.~L. Vit{\'o}ria, M.~V. d.~S. Silva and
  H.~Belich, \emph{{Light deflection and gravitational lensing effects inspired
  by loop quantum gravity}},
  \href{http://dx.doi.org/10.1088/1475-7516/2025/06/034}{\emph{JCAP} {\bf 06}
  (2025) 034}, [\href{http://arxiv.org/abs/2503.06373}{{\tt 2503.06373}}].

\bibitem{Donoghue:2014yha}
J.~F. Donoghue and B.~K. El-Menoufi, \emph{{Nonlocal quantum effects in
  cosmology: Quantum memory, nonlocal FLRW equations, and singularity
  avoidance}}, \href{http://dx.doi.org/10.1103/PhysRevD.89.104062}{\emph{Phys.
  Rev. D} {\bf 89} (2014) 104062}, [\href{http://arxiv.org/abs/1402.3252}{{\tt
  1402.3252}}].

\bibitem{El-Menoufi:2015cqw}
B.~K. El-Menoufi, \emph{{Quantum gravity of Kerr-Schild spacetimes and the
  logarithmic correction to Schwarzschild black hole entropy}},
  \href{http://dx.doi.org/10.1007/JHEP05(2016)035}{\emph{JHEP} {\bf 05} (2016)
  035}, [\href{http://arxiv.org/abs/1511.08816}{{\tt 1511.08816}}].

\bibitem{Birrell:1982ix}
N.~D. Birrell and P.~C.~W. Davies, \emph{{Quantum Fields in Curved Space}}.
\newblock Cambridge Monographs on Mathematical Physics. Cambridge University
  Press, Cambridge, UK, 1982.
\newblock 10.1017/CBO9780511622632.

\bibitem{Buchbinder:1992rb}
I.~L. Buchbinder, S.~D. Odintsov and I.~L. Shapiro, \emph{{Effective action in
  quantum gravity}}.
\newblock 1992.

\bibitem{Donoghue:2015nba}
J.~F. Donoghue and B.~K. El-Menoufi, \emph{{Covariant non-local action for
  massless QED and the curvature expansion}},
  \href{http://dx.doi.org/10.1007/JHEP10(2015)044}{\emph{JHEP} {\bf 10} (2015)
  044}, [\href{http://arxiv.org/abs/1507.06321}{{\tt 1507.06321}}].

\bibitem{Wald:1984rg}
R.~M. Wald, \emph{{General Relativity}}.
\newblock Chicago Univ. Pr., Chicago, USA, 1984.
\newblock 10.7208/chicago/9780226870373.001.0001.

\bibitem{Delgado:2022pcc}
R.~C. Delgado, \emph{{Quantum gravitational corrections to the entropy of a
  Reissner\textendash{}Nordstr\"om black hole}},
  \href{http://dx.doi.org/10.1140/epjc/s10052-022-10232-0}{\emph{Eur. Phys. J.
  C} {\bf 82} (2022) 272}, [\href{http://arxiv.org/abs/2201.08293}{{\tt
  2201.08293}}].

\bibitem{Penrose:1964wq}
R.~Penrose, \emph{{Gravitational collapse and space-time singularities}},
  \href{http://dx.doi.org/10.1103/PhysRevLett.14.57}{\emph{Phys. Rev. Lett.}
  {\bf 14} (1965) 57--59}.

\bibitem{Penrose:1969pc}
R.~Penrose, \emph{{Gravitational collapse: The role of general relativity}},
  \href{http://dx.doi.org/10.1023/A:1016578408204}{\emph{Riv. Nuovo Cim.} {\bf
  1} (1969) 252--276}.

\bibitem{Poisson:1989zz}
E.~Poisson and W.~Israel, \emph{{Inner-horizon instability and mass inflation
  in black holes}},
  \href{http://dx.doi.org/10.1103/PhysRevLett.63.1663}{\emph{Phys. Rev. Lett.}
  {\bf 63} (1989) 1663--1666}.

\bibitem{Poisson:1990eh}
E.~Poisson and W.~Israel, \emph{{Internal structure of black holes}},
  \href{http://dx.doi.org/10.1103/PhysRevD.41.1796}{\emph{Phys. Rev. D} {\bf
  41} (1990) 1796--1809}.

\bibitem{Ori:1991zz}
A.~Ori, \emph{{Inner structure of a charged black hole: An exact mass-inflation
  solution}}, \href{http://dx.doi.org/10.1103/PhysRevLett.67.789}{\emph{Phys.
  Rev. Lett.} {\bf 67} (1991) 789--792}.

\bibitem{Bonanno:2025bgc}
A.~Bonanno, A.~Panassiti and F.~Saueressig, \emph{{Cauchy Horizon (In)Stability
  of Regular Black Holes}},  \href{http://arxiv.org/abs/2507.03581}{{\tt
  2507.03581}}.

\bibitem{Misner:1973prb}
C.~W. Misner, K.~S. Thorne and J.~A. Wheeler, \emph{{Gravitation}}.
\newblock W. H. Freeman, San Francisco, 1973.

\bibitem{Claudel:2000yi}
C.-M. Claudel, K.~S. Virbhadra and G.~F.~R. Ellis, \emph{{The Geometry of
  photon surfaces}}, \href{http://dx.doi.org/10.1063/1.1308507}{\emph{J. Math.
  Phys.} {\bf 42} (2001) 818--838},
  [\href{http://arxiv.org/abs/gr-qc/0005050}{{\tt gr-qc/0005050}}].

\bibitem{Weinberg:1972kfs}
S.~Weinberg, \emph{{Gravitation and Cosmology}: {Principles and Applications of
  the General Theory of Relativity}}.
\newblock John Wiley and Sons, New York, 1972.

\bibitem{Liu:2024wal}
H.~Liu, M.-Y. Lai, X.-Y. Pan, H.~Huang and D.-C. Zou, \emph{{Gravitational
  lensing effect of black holes in effective quantum gravity}},
  \href{http://dx.doi.org/10.1103/PhysRevD.110.104039}{\emph{Phys. Rev. D} {\bf
  110} (2024) 104039}, [\href{http://arxiv.org/abs/2408.11603}{{\tt
  2408.11603}}].

\bibitem{EventHorizonTelescope:2022urf}
{\scshape Event Horizon Telescope} collaboration, K.~Akiyama et~al.,
  \emph{{First Sagittarius A* Event Horizon Telescope Results. V. Testing
  Astrophysical Models of the Galactic Center Black Hole}},
  \href{http://dx.doi.org/10.3847/2041-8213/ac6672}{\emph{Astrophys. J. Lett.}
  {\bf 930} (2022) L16}, [\href{http://arxiv.org/abs/2311.09478}{{\tt
  2311.09478}}].

\bibitem{Tiede:2022grp}
P.~Tiede, M.~D. Johnson, D.~W. Pesce, D.~C.~M. Palumbo, D.~O. Chang and
  P.~Galison, \emph{{Measuring Photon Rings with the ngEHT}},
  \href{http://dx.doi.org/10.3390/galaxies10060111}{\emph{Galaxies} {\bf 10}
  (2022) 111}, [\href{http://arxiv.org/abs/2210.13498}{{\tt 2210.13498}}].

\bibitem{Abe:2010ap}
F.~Abe, \emph{{Gravitational Microlensing by the Ellis Wormhole}},
  \href{http://dx.doi.org/10.1088/0004-637X/725/1/787}{\emph{Astrophys. J.}
  {\bf 725} (2010) 787--793}, [\href{http://arxiv.org/abs/1009.6084}{{\tt
  1009.6084}}].

\bibitem{EventHorizonTelescope:2019ggy}
{\scshape Event Horizon Telescope} collaboration, K.~Akiyama et~al.,
  \emph{{First M87 Event Horizon Telescope Results. VI. The Shadow and Mass of
  the Central Black Hole}},
  \href{http://dx.doi.org/10.3847/2041-8213/ab1141}{\emph{Astrophys. J. Lett.}
  {\bf 875} (2019) L6}, [\href{http://arxiv.org/abs/1906.11243}{{\tt
  1906.11243}}].

\bibitem{EventHorizonTelescope:2025vum}
{\scshape Event Horizon Telescope} collaboration, K.~Akiyama et~al.,
  \emph{{Horizon-scale variability of M87* from 2017--2021 EHT observations}},
  \href{http://arxiv.org/abs/2509.24593}{{\tt 2509.24593}}.

\bibitem{LSST:2008ijt}
{\scshape LSST} collaboration, {\v{Z}}.~Ivezi{\'c} et~al., \emph{{LSST: from
  Science Drivers to Reference Design and Anticipated Data Products}},
  \href{http://dx.doi.org/10.3847/1538-4357/ab042c}{\emph{Astrophys. J.} {\bf
  873} (2019) 111}, [\href{http://arxiv.org/abs/0805.2366}{{\tt 0805.2366}}].

\bibitem{Euclid:2024yrr}
{\scshape Euclid} collaboration, Y.~Mellier et~al., \emph{{Euclid. I. Overview
  of the Euclid mission}},
  \href{http://dx.doi.org/10.1051/0004-6361/202450810}{\emph{Astron.
  Astrophys.} {\bf 697} (2025) A1},
  [\href{http://arxiv.org/abs/2405.13491}{{\tt 2405.13491}}].

\end{thebibliography}\endgroup

\end{document}